# SWS- promoting MHb-IPN-MRN circuit opposes the theta- promoting circuit, active wake and REM sleep.

## A circuit-based model of functional properties.

Karin Vadovičová

ABSTRACT Hippocampus was connected to medial habenula (MHb) by multisynaptic axonal tracts in my previous DTI study. These probabilistic tracts linked hippocampus to septum, and amygdala to bed nucleus of stria terminalis (BNST). The axons from septum and BNST passed by anteromedial thalamic nucleus (AM) to MHb, then from MHb to pineal gland, known for control of circadian cycles and sleep. Question is what is the MHb doing and why it receives information from hippocampus via septum? So this study explores the connectivity of MHb, predicts its functional role and how it is linked to memory replay. My combination of known findings about septum and MHb connectivity and function led to this circuit-based idea/hypothesis that posterior septum activates MHb, MHb activates interpeduncular nucleus (IPN), and then IPN stimulates MRN and its serotonin release. Proposed idea is that this MHb-IPN-MRN circuit promotes slow wave sleep (SWS), high serotonin and low acetylcholine state. My prediction is that this SWS-promoting circuit reciprocally suppresses the theta oscillations promoting circuit, linked to high acetylcholine levels in brain, and formed by supramamillary area (SUM) projections to the medial septum (MS) that induces theta rhythm in hippocampus and other theta-coupled regions. The MHb-IPN-MRN pathway likely inhibits, possibly reciprocally, also some regions that have stimulating input to the theta-generating SUM and MS, such as the wakefulness promoting nucleus incertus (NI), posterior hypothalamus (PH), lateral hypothalamus (LH) and laterodorsal tegmentum (LDT), as well as the REM sleep inducing neurons in LDT and reticular nucleus pontis oralis (PnO). Thus, proposed SWS- promoting circuit attenuates the output of the theta-promoting regions, both the active wake-on and REM-on regions. The theta rhythm in wake state is linked to recording and binding information with their spatio-temporal and relational context by hippocampus, while the SWS supports rest, replay and transfer of hippocampally stored inter-connected information/memory traces and their cortical reactivations, e. g. in retrosplenial cortex linked to autobiographic memory or in prefrontal cortex that can combine information from any sources.

## Introduction. Known role of serotonin in inducing SWS.

Serotonin signal in the brain induces behavioural and emotional relaxation, slow-down and rest while dopamine promotes action, locomotion, motivation, speeding up. Serotonin is present early during embryogenesis, probably to slow down/lower heart rate and blood pressure (by activating parasympathetic system). Later on the same serotonin slows down skeletal muscles movement (running, locomotion) and breathing, attenuates sympathetic system by inhibiting amygdala and dorsal anterior cingulated cortex (dACC), plus might enable the relaxation phase in ON-OFF rhythmic motor activity of axial muscles in fish where contraction alternates with extension. Brain serotonin slows down locomotion, calms down the active avoidance, negative emotions and drive, makes us rest and relax, promotes well-being (Vadovičová and Gasparotti, 2014), satisfaction and reconstructive (deep) stage of sleep called slow wave sleep or non-REM (SWS or NREM) sleep. SWS shows high amplitude, low frequency EEG, while the rapid eye movement (REM) sleep shows low amplitude, high frequency EEG oscillations in cortex (Jouvet, 1962). The SWS is known for its recovery-restorative function, calcium cascade of nocturnal proteosyntesis and also for the replay and transfer of the contextually/relationally bound information from the hippocampus to cortex, by replaying the waking neuronal firing patterns in hippocampus and cortex during temporally coupled hippocampal sharp-wave/ripples (SWRs) and cortical spindles (Buzsáki, 1989). The 5-10 Hz oscillations in the hippocampal local field potential called theta rhythm occur during voluntary exploration, locomotion and REM sleep (O'Keefe and Recce, 1993; Bland and Oddie, 2001; Buzsaki, 2002). REM sleep is proportionally more abundant in young mammals (Roffwarg et al. 1966). The EEG patterns of sleep and wakefulness become adult-like only after full development of the cortex (Blumberg et al., 2003). The fast EEG synchrony typical of wakefulness develops through adolescence in humans, similar to prolonged cortical maturation in higher primates (Uhlhaas et al., 2009).

Serotonin from the dorsal raphe nucleus (DRN) has been functionally grouped with the arousal-promoting neuromodulators important for wake state: noradrenaline, histamine, orexin (hypocretin) and acetylcholine. Orexin stabilizes wakefulness, as its deficiency causes narcolepsy disorder with inability to maintain long waking periods, with abrupt transitions into SWS sleep, and intrusions of REM sleep into waking (Sakurai, 2007). In addition, the mice deficient in norepinephrine or histamin sleep more and have no trouble to fall asleep after mild stress. In contrary, mice lacking serotonin receptor 5-HT1A (inhibitory receptor, also MRN autoreceptor) have 400% more REM sleep. The orexin, histamine and noradrenaline neurons have highest firing rate in wakefulness (Aston-Jones and Bloom, 1981). Extracellular electrophysiological recordings in freely moving cats have shown that DRN serotonergic neurons fire tonically during wakefulness, decrease their activity during slow wave sleep (SWS), and are nearly quiescent during REM (McGinty and Harper, 1976; Trulson and Jacobs, 1979). In contrary, the acetylcholine levels in the rat thalamus (innervated by mesopontine cholinergic neurons) are intermediate in SWS but high in both REM and wake-state (Williams et al., 1994). In addition the acetylcholine release during active waking is increased by approximately 75% compared to quiet waking (Marrosu et al., 1995). Interestingly the tonic dopamine firing seems to be present in all sleep-wake cycles, and burst firing was found in both wake and REM sleep (Dahan et al., 2007).

Raphe system lesion in cats caused permanent insomnia and diminution of cerebral serotonin (Jouvet et al., 1967) and MRN lesions produced uninterrupted theta (Maru et al., 1979; Yamamoto et al., 1979). Serotonergic firing is inversely correlated with the occurrence of ponto-geniculo-occipital PGO-waves (Lydic et al., 1983) that are typical for REM sleep. Inhibition of serotonin synthesis in cats induced PGO- waves also in waking state (Jacobs et al., 1972). Injection of serotonin precursor induced SWS and suppressed REM sleep for 5-6 hours (Bogdanski et al., 1958; Monnier and Tissor, 1958; Costa et al, 1960; Delorme, 1966). So these SWS-promoting effects of serotonin are not caused by the wakefulness linked DRN and its cortical, striatal and amygdalar projections involved in choice behavior and well-being, but by the MRN that has the needed subcortical afferents and efferents to interact with the other SWS-promoting regions, and to suppress the theta, wakefulness and REM sleep promoting regions. Main MRN efferents target midline forebrain including supramamillary nucleus (SUM), medial mamillary body (MMB), medial septum and vertical diagonal band of Broca (MS/vDBB), posterior hypothalamus (PH), lateral habenula (LHb), perifornical hypothalamus (PeF), midline and intralaminar thalamus, zona incerta (ZI) and hippocampus

(Vertes and Linley, 2008). Main MRN and DRN afferents are from lateral and medial preoptic area (LPO, MPO), LHb, lateral hypothalamus (LH) and PeF (both produce orexin), dorsomedial nuclei of hypothalamus (DMN), midbrain nuclei, locus coeruleus (LC), caudal raphe nuclei, laterodorsal thalamus (LDT) and periacqueductal grey (PAG). Additional MRN afferents are from IPN and MB, while the DRN afferents are from amygdala, bed nucleus of stria terminalis (BNST), lateral septum (LS), substantia nigra, DBB and tuberomamillary nucleus (TMN). Serotonin hyperpolarizes cholinergic burst neurons in the rat LDT in vitro (Luebke et al., 1992), so opposes/suppresses the theta circuit, wake and REM sleep states that require this acetylcholine signal. Serotonergic MRN projections inhibit also theta bursting of MS/vDBB, the source of hippocampal theta rhythm (Kinney et al., 1996). This was found by recordings in mice MS/DBB and hippocampus after inhibition of MRN by 5-HT1A agonist on its autoreceptors. This is a functional evidence for known antagonism between the SWS-provoking MRN and the theta-generating MS/vDBB region.

Cholinergic agonists promote theta rhythm and PGO waves in rats and cats, but serotonergic neurons inhibit theta rhythm and PGO waves. The reciprocal interaction model of sleep control (Hobson and McCarley 1975; McCarley and Hobson 1975) proposed that REM-on neurons increase their firing just prior and during REM, while REM-off neurons show reverse pattern. The cholinergic REM-on neurons were located in LDT and PPT. The norepinephrine and serotonin REM-off neurons (from locus coeruleus and raphe nucleus) were found to be inhibited during REM sleep, and to inhibit REM-on (cholinergic) neurons in LDT/PPT during waking and SWS. Many studies support this mutual inhibition between the REM sleep and SWS circuits. For example reversible inactivation of LC and DRN by cooling decreased REM sleep (Cespuglio et al, 1982). In cats were REM-on LDT/PPT neurons inhibited by serotonin agonist, while WAKE-on/REM-on neurons were unaffected (Thakkar et al., 1998). Role of acetylcholine in promoting REM sleep in animals was found in early studies (George et al., 1964; Hernandez-peon et al., 1964). Multiple studies in humans showed that enhancement of cholinergic tone decreases REM sleep latency and increases its amount (Berkowitz et al. 1990; Hohagen et al., 1993; Lauriello et al, 1993; Riemann et al., 1994; Sitaram and Gillin, 1980; Sitaram et al., 1976).

Based on combination of functional findings and anatomical connectivity of the medial habenula (MHb), interpeduncular nucleus (IPN) and serotonergic MRN in the literature, this new hypothesis shows how the MHb → IPN → MRN circuit (linked to the low acetylcholine and high serotonin state) promotes slow wave sleep. This idea proposes antagonism (reciprocal inhibition) between this SWS-promoting circuit and the theta- promoting circuit formed by SUM and MS/vDBB. The strength of theta oscillations increases during both active wake and REM sleep. Theta rhythm is linked to high acetylcholine release in brainstem and thalamus. This model predicts also a functional opposition between the MHb → IPN → MRN circuit and some regions that stimulate the theta-promoting circuit: nucleus incertus (NI), PH, LH, LDT and ventral tegmental nucleus of Gudden (VTg). Both MRN and LPO are known to promote SWS. As the IPN projects to LPO, it might also potentiate the LPO effects.

After first describing the function and connectivity of proposed MHb-IPN-MRN circuit, the following sections will try to show its interesting known and predicted interactions with several regions involved in control of wakefulness, sleep, in theta oscillations, or contextual memory formation and recall.

The medial habenula pathway.

This study extends the prevailing models of sleep and wakefulness control in the brain, by adding the MHb and IPN regions to the known systems and circuits. It will provide wider context, circuit based and functional evidence for the idea that posterior septum (PS) activates MHb, which activates IPN, which stimulates serotonergic MRN to promote SWS and to inhibit the theta-, wakefulness and REM sleep promoting regions (Vadovičova, 2014). This idea came from my DTI probabilistic tractography study in humans, which showed disynaptic axonal tracts that connected hippocampus and amygdala (via dorsal thalamus and via fornix) to septum and anterior BNST, which both projected to MHb. The observed axonal tracts from hippocampus and amygdala passed tightly by both septum and aBNST, then turned posteriorly by anteromedial thalamus (AM), and reached to MHb via stria medialis, then projected from MHb to the pineal gland. This pathway resembled the MHb afferents known from tracing studies in rat. The observed tract branched around septum and BNST to reach also hypothalamus, and via the amygdalofugal pathway passing by/through substantia innominata (SI) it was linked also with dorsal central amygdala (CeA) and hippocampus. Also the vACC showed axonal links to (or from) septum and BNST, so the well-being signal from vACC might control strength of septal and BNST output. Now we need to look into anatomy and physiology data to see what these diverse septo-habenular projections could do, to guess their possible functions and roles. Anatomical studies show input to MHb from posterior septum and BNST, and main output from MHb to IPN, which then projects to MRN. So it is possible that these projections have stimulating effects that lead to serotonin release from MRN.

### Medial habenula, MHb. Its connectivity and proposed role.

The main MHb input comes from supracommissural septum (Herkenham and Nauta, 1977). Triangular septum (TS) with input from hippocampal dentate gyrus; and septofimbrial nucleus (SF) with input from fimbria of hippocampus form posterior septum (PS) and project to MHb (Raisman, 1966). The MHb receives input from the glutamatergic and ATPergic triangular septum, from cholinergic septofimbral nuclei, from GABAergic medial septum and vertical DBB (MS/vDBB), from serotonergic raphe, noradrenergic locus coeruleus neurons and sympathetic superior cervical ganglion, from substance P-ergic anterior medial part of BNST (amBNST or BAC, bed nucleus of anterior commissure), and from dopaminergic ventral tegmental area,VTA (Herkenham and Nauta, 1977; Gottesfeld, 1983; Qin and Luo, 2009; Yamaguchi et al., 2013). MHb has also afferents from LDT, and reciprocal connections with MS/vDBB (Woolf, 1991), which my model suggest to have all inhibitory effects. Model further proposes that triangular and septofimbrial septal nuclei stimulate MHb, while the MS/vDBB, and noradrenergic LC and sympathetic system neurons attenuate the output of MHb. The theta-generating MS/DBB likely inhibits MHb because MHb via IPN stimulates serotonergic MRN that promotes SWS and suppresses few regions of theta- promoting circuit: SUM, MS/vDBB and LDT. Similarly the stress (and light) evoked noradrenaline (NA) input suppresses MHb pathway to interrupt or postpone induction of SWS in the unsafe circumstances. In addition the substance P likely activates MHb when the circumstances are safe enough to afford sleep (without imminent threat to survival). This might be supported by the TS and aBNST/BAC connectivity that forms two parallel pathways: dentate gyrus → TS → ventral MHb → IPN core, and the medial amygdala (MeA) → BAC → dorsal MHb → lateral IPN (Yamaguchi et al., 2013) and by the increased fear and arousal after BAC lesions. In addition the BAC input comes from the anxiety decreasing medial amygdala (MeA). In my opinion the amBNST/BAC projections might also attenuate SUM, LC, VTA, SI, paraventricular thalamus (PVT), periacqueductal grey (PAG), CeA, and stimulate besides the MHb also the LPO.

Medial habenula projects to and activates pineal population of silent cells (Axelrod, 1970; Ronnekleiv and Moller, 1979, Ronnekleiv et al., 1980) known to produce melatonin in the dark period. Light stimulation on retina inhibits melatonin synthesis by activating suprachiasmatic nucleus (SCN), which then activates paraventricular hypothalamic nucleus (PVN, known for hormonal reaction to stress) that stimulates sympathetic system in superior cervical ganglia that inhibits pineal gland and MHb by norepinephrine release (Teclemariam-Mesbah et al. 1999). Some PVN fibres innervate pineal gland directly (Reuss and Moller, 1986). Norepinephrine (NE) released from sympathetic terminals at pineal gland attenuates nicotinic cholinergic input from parasympathetic system and MHb (Yoon et al., 2014). Parasympathetic autonomic system is linked to slowing down heart beat and respiration, to rest and digestion.

The superior MHb is glutamatergic and also expresses Interleukin-18 (IL-18); the dorsal MHb is both glutamatergic and substance P-ergic. Both the superior and dorsal MHb have hight density of mu-opioid receptor; while the inferior parts of MHb are both cholinergic and glutamatergic (Sugama et al, 2002; Aizawa et al., 2012). MHb sends topographically organized glutamate, substance P and acetylcholine output to IPN (Qin and Luo, 2009; Ren et al., 2011; Herkenham and Nauta, 1979). So it is interesting that MHb that belongs to the SWS- and serotonin- promoting system produces also acetylcholine (the main neuromodulator of the opposing theta- promoting circuit). Substance P from the MHb was found released in lateral habenula (Kim and Chang, 2005; Antolin-Fontes et al., 2015) and VTA (Claudio Cuello et al., 1978). My prediction is that MHb attenuates both LHb and VTA output to enable SWS. The MHb projects to IPN via the internal part of fasciculus retroflexus and via IPN it controls MRN, LPO and LDT (Herkenham and Nauta, 1979; Groenewegen et al., 1986). So the IPN likely stimulates MRN and LPO, and inhibits LDT.

Based on its anatomical connectivity and interactions with neuromodulators, this model proposes that MHb stimulates SWS via the MHb → IPN → MRN pathway that also suppresses the theta, wakefulness, alertness and REM driving regions. This is supported by dense mu opioid receptors (that bind morphine, the sleep inducing substance) in MHb, and by circadian rhythmicity of MHb neurons (McCormick and Prince, 1987; Quick et al., 1999; Guilding and Piggins, 2007). By the markedly increased MHb and IPN metabolic activity during anesthesia (pentobarbital, ether and chloral hydrate) in rats (Herkenham, 1981), and also by the fact that the MHb neurons produce sleep promoting interleukin IL-18 (Sugama et al., 2002) and might control via IPN the MRN serotonin. Further evidence comes from high firing rates of MRN cells during SWS and during non-exploratory waking states in rats (when not recording new information by hippocampus), and their low firing rates during the theta linked exploration and REM sleep (Jacobs and Azmitia, 1992; Marrosu et al., 1996). Firing of serotonergic DRN neurons is high in wakefulness, lower in SWS and minimal in REM sleep (Saper et al., 2001). The cholinergic activity in LDT/PPT as well as the acetylcholine levels in cortex and hippocampus are high in wakefulness and REM sleep (Sakai, 1980; Marrosu et al., 1995). Discharge rate of noradrenergic LC neurons is highest during active waking, significantly lower during quiet waking, and ceased during SWS and REM sleep (Takahashi et al., 2010). Another studies found lower LC firing during SWS in rats than during wakefulness (Aston-Jones and Bloom, 1981). The lack or low NE in SWS fits its alarm/alert inducing function. Histamine neurons of tuberomamillary nucleus (TMN) in posterior hypothalamus (PH) are active only in wakefulness, highest at high vigilance, low at quiet waking and silent during SWS and REM (Takahashi et al., 2006).

Wake, SWS and REM circuits (ON and OFF regions).

The LPO or ventrolateral preoptic area (VLPO) induces SWS by inhibiting cholinergic LDT/PPT and nucleus basalis of Meynert NBM (source of cortical stimulation and gamma coupling, inhibited by adenosine), noradrenergic LC, histaminergic TMN (and PH), orexinergic LH and serotonergic DRN, so by inhibiting the wake-promoting monoaminergic arousal system (Strecker et al., 2000). Reciprocally the TMN, LC, orexinergic LH and GABAergic NBM inhibit VLPO, either directly or via interneurons (Steininger et al., 2001). LDT/PPT and orexinergic LH stimulate cholinergic and parvalbumin containing NBM neurons that induce fast gamma coupling in cortex (Kim et al., 2015). Orexin stabilizes wakefulness by exciting cortex, TMN, LC, DRN, VTA and LDT/PPT (Kilduff and Peyron, 2000; Saper et al., 2001, Bernard et al., 2002).

In addition the serotonergic and noradrenergic (REM-off) regions are suppressed by REM-promoting PnO, SubCoeruleaus (SubC) or medullar dorsal paragigantocerullar nuclei (DPGi), directly or via nearby GABAergic REM-on neurons (Gervasoni et al, 1998; Gervasoni, 2000). Reciprocally, the LC and DRN inactivation increased REM sleep (Cespuglio et al., 1982). Norepinephrine inhibited mesopontine cholinergic (probably REM-on) neurons (Williams and Reiner, 1993), while serotonin injection into SubCoeruleus in rats suppressed PGO waves in pons without affecting thalamic or cortical PGO (Datta et al., 2003). REM-on LDT/PPT neurons were inhibited by 5-HT 1A agonist in cats while the wake-on/REM-on neurons of LDT/PPT were not (Thakkar et al., 1998). Also the LPO that induces SWS is inhibited by REM-on nuclei that stimulate GABA (REM-on) neurons nearby LPO. Both the SWS-on, deep mesencephalic reticular nucleus (also named LPT) and the wake-on, vlPAG nucleus (that stimulates locomotion and motor neurons) do inhibit REM-on regions SubC and PnO/RPO. The PnO is known to be stimulated by REM-on neurons of PPT/LDT.

So my model predicts reciprocally inhibitory interactions between the SWS- promoting MHb → IPN → MRN circuit and the theta-promoting circuit SUM → MS/vDBB → hippocampus. It also proposes the opposition (reciprocal inhibition) between this MHb pathway activation and the activity of regions that stimulate the theta-generating circuit: NI, VTg, LDT and PH; as well as those regions that stimulate wakefulness: histaminergic TMN, orexinergic LH, gamma coupling inducing NBM, value-signaling/action-urging VTA, and alert/alarm linked LC. So there is also functional opposition (mutual inhibition) between the SWS- promoting circuit (MHb-IPN-MRN and LPO) versus the wakefulness promoting orexin, histamine, norepinephrine and the locomotion triggering and value-signaling dopamine system. SWS-promoting circuit might show similar opposition with regions of the REM-promoting circuits, REM-on LDT/PPT, PnO, PnC or SubC.

Brain regions with theta rhythm.

The theta oscillations in hippocampus occur during awake exploration and during REM sleep (Vanderwolf, 1969). They are generated by supramamillary nuclei (SUM) that determine theta frequency and activate medial septum and vertical limb of diagonal band of Broca (MS/vDBB) that determine theta amplitude and induce theta oscillations in hippocampus and other target regions that show theta-rhythm (Pan and McNaughton, 2004). The SUM is activated by nucleus pontis oralis PnO/RPO in anesthetized rats during REM (Vertes and Martin, 1988; Vertes and Kocsis, 1997; Oddie et al., 1994), possibly also by LDT. During waking state is SUM activated for example by posterior hypothalamus (PH), orexinergic LH, dopaminergic (VTA) and nucleus incertus (NI) input, while the NI induces theta rhythm in

MS by release of insulin-like peptide relaxin-3 (Ma et al., 2009). Another way to induce theta coupling in the brain might be via reciprocal connections between prefrontal cortex and SUM. Both SUM and MS project to hippocampus and medial mamillary body (MMB). The MMB and SUM have reciprocal connections with the ventral tegmental nucleus of Gudden (VTg) that (similar to SUM, MMB and MS) shows theta oscillations and similar to NBM and hDBB contains neurons with parvalbumin. Parvalbumine GABAergic cells from hDBB and NBM generate fast gamma synchronization in cortex (so possibly also in VTg). Medial part of MMB projects to the anteromedial nuclei (AM), while its lateral part projects to the anteroventral (AV) nuclei of ATN. Both AV and AM show theta rhythm. Prefrontal cortex projects to AV (top down control) but is reciprocally connected with AM, so AM might contextually bias the source of information in prefrontal cortex working memory depending on event-based context and memories and associations linked to it. Prefrontal cortex involved in goal-directed control of behavior, evaluation, predictions, decision making and planning projects also to MMB. Hippocampal efferents from subiculum project via fornix to septum (via fornix but my DTI results showed also dorsomedial thalamic tract), anterior thalamic nuclei and mamillary bodies (Aggleton et al., 2005). Further regions that demonstrate theta rhythm are subicular, retrosplenial, entorhinal, perirhinal, posterior cingulate cortex (PCC), ACC and medial prefrontal cortices that receive afferents from MS/DBB, anteromedial and anteroventral nuclei of ATN. Based on its connectivity, my prediction is that AM has role in automatic context- based selection of information from episodic and relational memories that will reach prefrontal cortex and working memory based on their associations with ongoing events. Prefrontal cortex input to AM and AV might on the other hand help to select recall of those episodic memories, events, scenarios, cognitive schemes and task rules that are relevant to current context, goals, task set, situation, and thoughts in working memory.

Interestingly both the theta inducing MS/vDBB and the gamma inducing hDBB have reciprocal connections (Woolf, 1991) with the IPN, MHb, LHb, hippocampus, amygdala, subiculum, entorhinal cortex, piriform cortex, retrosplenial cortex, cingulate cortex (perhaps both PCC and ACC), insula (codes aversive properties of things and subjects, pain and taste), and temporal pole (codes identities of objects and subjects). Thus, this model predicts reciprocal inhibition of the MS/vDBB and hDBB by MHb and IPN, as well as the inhibition of MS/vDBB and hDBB by LHb (to decrease theta and gamma rhythm linked to recording of new information and arousal). These cortical regions and amygdala might reciprocally induce the MS/vDBB and hDBB firing and be driven by them into theta and gamma coupling. The theta coupling in cortex might also be induced by SUM. Other afferents of MS/DBB are from PPT/LDT, LC, VTA, orexinergic LH, DRN and MRN (Woolf, 1991).

Supramamillary nucleus, SUM and some of its interactions.

SUM controls the frequency of hippocampal theta activity via MS/DBB, while MS/vDBB controls its amplitude (Pan and McNaughton, 2004). SUM is known to stimulate theta rhythm during exploration in rats (Vertes and Kocsis, 1997) that gets disrupted by serotonergic MRN. SUM in rats has reciprocal connection with MMB, VTg, PH, LH, AH, LDT, MS/DBB, lateral septum (LS), PAG, LPO, medial preoptic hypothalamic nucleus (MPO, involved in regulation of body temperature), MRN, DRN, VTA (VTA encodes expected reward value signal and novelty), cognitive anterior cingulate cortex (homologue of human cognitive PFC), medial and ventrolateral orbital cortex (processing of rewards and punishments). SUM afferents come also from IPN, lateral habenula (LHb), (Kiss et al., 2002), VMH, BNST, subiculum and nucleus prepositus. SUM efferents project also to anteromedial and anteroventral thalamic nuclei (AM, AV), reuniens nuclei, intralaminar thalamic nuclei, centromedian thalamic nuclei (CM), mediodorsal thalamus (MDT, involved in value-based selection of choices that are further encoded in prefrontal working memory), substantia innominata (SI, which includes primate NBM homologue), hippocampus, dentate gyrus (receives strong input), entorhinal cortex (EC), frontal cortex, subthalamic nucleus (STN), amygdala, LC and cerebellar nuclei (CN), (Vertes, 1988, 1992; Hayakawa et al., 1993; Shibata, 1987; Risold and Swanson, 1997; Swanson, 1982; Thinschmidt, 1993; Kiss et al., 2002, Contestabile and Flumerfelt, 1981). So this model predicts stimulating input to SUM from wakefulness promoting regions of PH, LH, LDT (from wake-on/REM-on LDT cells), from VTA (with dopamine that signals value), from PAG, and prefrontal cortex (that informs SUM about meanings of things and events around us). Model proposes inhibitory input from LHb, and from SWS- promoting IPN, MRN and LPO to SUM. Similarly, this model suggests stimulatory effects of SUM on VTg, AM, AV, CM, MDT, SI, amygdala, entorhinal and prefrontal cortex function, and predicts inhibitory effect of SUM efferents on MRN output.

Ventral tegmental nucleus of Gudden, VTg and its connectivity.

VTg has reciprocal projections with MMB. VTg has input from prefrontal, cingulate, insular and retrosplenial cortex, MRN, IPN and LHb, maybe even from NBM, PH, LH and LPO as VTg receives input from basal forebrain and hypothalamus (Irle et al., 1984). Both VTg and SUM receive LHb, IPN and MRN projections, so this model predicts that they have inhibitory effect on both these regions that demonstrate theta oscillations. Because the VTg is needed for alternate choice working memory and during mental navigation tasks, but not during SWS sleep. In addition the VTg has possibly stimulating input from dorsal tegmental nucleus of Gudden (DTg), vestibular nucleus (head movement signals), substantia nigra pars compacta (SNc, that signals informational value to brain, what is meaningful, relevant, what way of doing things is right), VTA (signals expected reward value), LC (alarm signal), PAG (fear response) and from dopaminergic cells in zona incerta (ZI, linked to locomotion). VTg has input also from fields of Forel, nucleus of Darkschewitsch (reflex gaze), interstitial nucleus of Cajal (integration of head and eye movements) and nucleus prepositus hypoglossi. Many VTg neurons projecting to MMB are parvalbumin positive, so capable to transfer fast gamma oscillations important for WM and focused attention.

Nucleus incertus, NI and its interaction with SWS-circuit.

The NI is known to induce theta rhythm and to increase spatial working memory performance (and theta-power in hippocampus), by releasing peptide relaxin-3 into MS/DBB (Ma et al, 2009). Even in the anaesthetised rats did the relaxin-3 antagonist in medial septum decreased the PnO-induced hippocampal theta power (Ma et al, 2009). The NI in rats has reciprocal connections with cortical affective evaluation regions: prelimbic cortex (homologue of dorsal anterior cingulated cortex, dACC in primates), medial and ventrolateral orbital cortex; with cognitive anterior cingulate cortex (cognitive PFC homologue); with retrosplenial cortex important for autobiographic memory; with subcortical MRN, IPN (Goto et al., 2001; Aizawa et al., 2012), LPO, ZI, medial LHb, SUM, PH, LH, ventrolateral PAG, MS/vDBB (GABAergic negative feedback) and pontine reticular nucleus. Medial LHb inhibits MRN and projects to intermediate part of the IPN (Wang and Aghajanian, 1977; 2015; Kim, 2009). This model predicts that the cortical input likely stimulates via NI the theta coupling to enhance the recording of episodes with affective and informative meanings to us. Based on the properties of SWS- promoting circuit MHb-IPN-MRN, this model predicts that LHb and the theta- opposing regions IPN and MRN do reciprocally inhibit the output of the theta- and arousal inducing NI, and that LHb also inhibits the IPN output to MRN.

The NI in rats and mice has receptors for CRH (corticotropin releasing hormone), orexin, MCH (melanin concentrating hormone, linked to sleep), oxytocin, serotonin 5-HT1A and ghrelin (Bittencourt and Sawchenko, 2000; Greco and Shiromani, 2001; Marcus et al., 2001; Saito et al., 2001; Vaccari et al., 1998; Mani et al, 2014; Miyamoto et al., 2008). So the serotonergic MRN likely inhibits NI during SWS, while the wakefulness linked agents activate NI to enforce arousal, theta coupling and the recording of new events in hippocampus. This model suggests that histaminergic and orexinergic neurons of PH/LH stimulate NI, as both also promote wakefulness, and that LPO inhibits NI, as LPO promotes SWS that opposes theta state and arousal. NI in rats projects also to all hippocampus especially to dorsal fimbria and ventral dentate gyrus, CA3 and subiculum, to entorhinal cortex, claustrum, SI, PPT, LS, triangular and septofimbrial septal nuclei, MHb, MMB, lateral mamillary body (LMB), AV, AM, AD, nucleus reuniens, CM, MDT , DRN, PVT, PVN, dorsomedial thalamic nucleus (DMN), VMN, MPO, supraoptic hypothalamic nucleus (SON), anterior hypothalamus (AH), arcuate nucleus, amygdala, BNST, contralateral NI, nucleus accumbens (NAc, involved in motivation/drive and inhibitory avoidance), ZI, VTA, SNc, and superior colliculus (SC), (Goto et al., 2001; Olucha-Bordonau et al., 2003 and 2012; Teruel-Marti et al., 2008). This model predicts that NI (besides its known effect on medial septum theta rhythm) activates its targets in PFC, RSC, hippocampus, BNST, LS, PAG, PVN, PVT, MDT and CM; but inhibits MHb, IPN, MRN and LPO leading to suppression of SWS.

Triangular and septofimbrial septal nuclei. Some of their interactions.

Triangular septum (TS) projects to MHb, IPN and LHb (Raisman, 1966). Posterior septum (PS) that includes TS and septofimbrial septal nuclei (SF) has GABAergic projections to NI (and NI projects to PS), SUM and MRN. Also medial septum and horizontal DBB (hDBB) have GABAergic projections to nucleus incertus (Sanchez-Perez et al., 2015). By proposing SWS- promoting role of PS, this model suggests that TS inhibits LHb to disinhibit the serootnergic MRN, and that TS and SF inhibit NI and SUM, but activate MRN (by inhibiting MRN interneurons). The NI is known to induce theta by its relaxin-3 release in MS (Ma et al., 2009). Horizontal DBB together with NBM is the main source of the cortical cholinergic innervation and the parvalbumin containing GABAergic projection neurons that induce fast gamma rhythm linked to thinking, attention, working memory and feature binding. So the hDBB might induce cholinergic activation or even gamma oscillations in NI towards informative and salient stimuli to increase temporal coupling (and temporal summation) towards important information.

Median raphe nucleus, MRN and its interactions.

Major MRN input is from IPN, LHb, MS, MB, LPO, MPO, LH, perifornical hypothalamus, DMN, PAG, LDT, LC and infralimbic cortex in rats (homologue of ventral ACC in humans). Medial LHb reciprocally inhibits serotonergic MRN (Wang and Aghajanian, 1977). MRN projects to medial mammillary body (MMB), SUM, PH, perifornical hypothalamus (PF/LH), cholinergic MS/vDBB, hDBB, NBM, septum, VTA, dopaminergic A13 region of ZI, LC, LDT/PPT, subcoeruleus, LHb, IPN, MPO, nucleus reuniens, mediodorsal thalamus (MDT), central medial, paracentral and central lateral nuclei of midline intralaminar thalamus, suprachiasmatic nucleus (SCN), hippocampus and NAc (Vertes and Martin, 1998, Vertes and Linley, 2008). Cortical projections from MRN are light and restricted to perirhinal, entorhinal and some prefrontal cortex. MRN stimulation enhances the secretion of gonadotropins (James et al., 1987). Glutamate input to MRN was increased during non-theta phase of anesthesia in rats but not after tail pinch (Varga et al., 1998). This evidence supports the proposed activation of MRN by MHb during SWS, as MHb is also stimulated by anesthetics, possibly disinhibited by morphine, and has indirect input to MRN via IPN. So this model predicted that MHb stimulates IPN, which stimulates MRN to release serotonin during SWS. Both LPO and MRN promote SWS and suppress some regions of cholinergic theta-promoting circuit. So this model proposes that LPO and possibly vACC activate MRN, while the MS/vDBB, MMB, LDT, orexinergic LH, noradrenergic LC, and fear response related PAG inhibit MRN. Similarly to the known MRN inhibition of SUM, this model suggests that MRN attenuates also the following theta- , working memory or arousal linked regions: MMB, MS/vDBB, histaminergic PH, orexinergic LH, VTA, A13 of ZI, intralaminar and midline thalamic nuclei. Hippocampal theta induced by SUM and MS activation was produced after MRN inhibition by glutamatergic antagonists, GABA agonist and serotonin 1A agonist (acting via MRN autoreceptor) in urethane anesthetized rats (Kinney et al., 1994; Kinney et al., 1995; Vertes et al, 1994, Kinney et al., 1996).

Interpeduncular nucleus, IPN

Main IPN input comes from MHb, less from medial LHb, MPO, ventral hypothalamus, MRN, DRN, dorsal and ventral tegmental nuclei of Gudden (DTg, VTg), LDT, PAG, SUM, premamillary nuclei, LC, sparse input comes from hDBB (Contestabile and Flumerfelt, 1981; Vertes and Fass, 1988), and some from VTA neurons that contain dopamine and corticotropin releasing factor (CRF), (Zhao-Shea et al., 2013). The IPN neurons are mostly GABAergic and project to MRN, DRN, LPO, MPO, LDT, DTg and VTg (strongly), medial mamillary body (MMB, weakly), NI, nucleus basalis of Meynert (BNM), LC, PAG, MS/DBB, LH, hypothalamus, entorhinal cortex (EC), hippocampus, MDT, nucleus gelatinosus, and midline thalamic nuclei (Groenewegen et al., 1986; Goto et al., 200; Vertes and Fass, 1988). LDT has reciprocal connections with IPN, but also with MRN and LPO. So it is possible that IPN, MRN and LPO attenuate LDT output, to promote SWS, and vice versa that LDT suppresses IPN, MRN and LPO.

As gabaergic parvalbumine containing NBM projection neurons induce cortical gamma (40-100 Hz) oscillations and wakefulness (Brown and McKenna, 2015) and project to VTg, while the VTg shows theta oscillations, also contains fast firing parvalbumin GABA neurons (projecting to MMb), and is needed in alternation task working memory (Dillingham et al., 2015), this model suggests an inhibitory effect of IPN on NBM and VTg. It proposes that IPN stimulates the SWS-promoting LPO and MRN serotonin release; but attenuates the output of theta and arousal promoting regions: VTg, DTg, MMB, NI, LDT, NBM, LC and fight-or-flight response of PAG. In addition the model predicts that medial LHb that inhibits MRN possibly inhibits also IPN (by targeting IPN interneurons) and itself is inhibited by MHb. The IPN interneurons might be targeted also by dopamine and CRH release from VTA to interrupt sleep. Similarly, the IPN is probably inhibited by DTg, VTg, LDT, PAG, hDBB, SUM, premamillary nuclei and LC.

Lateral preoptic area, LPO and some interactions predicted by model.

GABAergic neurons in the VLPO promote sleep by inhibiting arousal-promoting circuits, such as TMN, NBM, LC and DRN (Saper et al., 1997; Saper et al., 2001; Luppi et al., 1999; Lydic and Baghdoyan, 2005). LPO was found to promote SWS. LPO projects also to SUM, LDT and PPT, and based on their properties this model proposes that LPO inhibits them. This inhibition might be reciprocal. It was found that MRN serotonin inhibits LTD, PPT and SUM. Both LPO and LH project to LHb, RMTg and VTA.
So this model predicts stimulatory effect of LPO on rostromedial tegmental nucleus (RMTg), and inhibitory effect of LPO on LHb and VTA, to disinhibit the MRN serotonin release during SWS (suppressed by medial LHb, i.e. in deppression) but to inhibit the dopamine groups (SNc/A9, VTA/A10 and A11) that do facilitate locomotion, value-based learning, drive (go-for-it) and activate spinal motor neurons (A11). This claim is supported by robust locomotor activation after infusion of the GABAA agonist muscimol into the RMTg (Lavezzi et al., 2024). The orexinergic LH projections on the other hand likely stimulate VTA and inhibit LHb and RMTg. Another supporting fact is that the LDT is essential for burst firing of VTA dopamine neurons (Lodge DJ, Grace AA, 2006). Stimulation of (wake-on ?) PPT or LDT excites dopaminergic neurons of VTA and SNc (Lacey et al. 1990) and is needed for maintenance of burst firing of dopamine neurons (Lodge and Grace, 2006). The RMTg is known for reciprocal connections with LDT, PPT, pontine and medullary reticular formation, and for strongly suppressing effect on VTA, SNc and DRN (Perroti et al., 2005; Jhou et al., 2009 a, b; Kaufling et al., 2009; Balcita-Pedicino et al., 2011). This model proposes that RMTg mutually supresses LDT and PPT to decrease arousal and locomotion (e.g. after injuries, in sickness, in depression). So the LPO uses the LHb and RMTg as "switch off-switch on" buttons to enable SWS, by boosting the serotonergic MRN system and by suppressing the dopaminergic SNc and VTA system. Actually the LPO might activate just that part of RMTg (lateral ?) that suppresses SNc and VTA, because the serotonergic DRN firing is reduced during SWS but ceased only during REM sleep. Interestingly respiratory rate is higher in REM sleep compared with both non-REM sleep and wakefulness, in line with the inhibition/cessation of serotonergic firing at MRN and DRN, and with calming serotonergic effect on respiration.
Similarly to the LPO, also the SWS-on deep mesencephalic nuclei might suppress dopamine system. That population of BNST neurons, which promotes reactivity to threat, activates VTA and inhibits LPO, most probably inhibits also the LHb and RMTg to disinhibit motor reactivity. Also the MS/DBB has input to LHb, RMTg and VTA, possibly to induce there the theta coupling. This model predicts inhibitory effect of LPO projections to SUM, to suppress theta oscillations during SWS. The main inhibitory input to LPO comes from GABAergic neurons of NBM, lateral septum and BNST (Zahm et al., 1999, 2013; Zahm, 2006). Their role in fast gamma coupling, panic and anxiety response supports the LPO role in SWS. Threat suppresses SWS by inhibiting LPO, what leads to lower RMTg activity: causing rise in dopamine firing, agitation and impulsive response. Interestingly cholinergic projections from the mesopontine tegmentum inhibit the RMTg at muscarinic M4 receptors, while exciting VTA dopamine neurons at M5 receptors (Wasserman et al. 2013, 2014). So this acetylcholine might come from wake promoting LDT to disinhibit dopamine system, to fuel motion, motivated behaviour, drive, reward and value/meaning based learning in wakefulness and awareness. The LPO projects also to reticular thalamic nucleus involved in up and down states during SWS, and to hippocampus, cortex and parabrachial nucleus (PB, part of pain pathway), possibly to promote SWS sleep.

Lateral habenula, LHb and its proposed role in theta control and REM sleep

LHb directly and via RMTG activation strongly suppresses (for few milliseconds) dopamine and serotonin release (Christoph et al., 1986; Wang and Aghajanian, 1977; Park, 1987; Ji and Shepard, 2007; Matsumoto and Hikosaka, 2007). Looking at the lateral habenula connectivity and function my conclusion is that LHb might be used as a "switch-off" button to suppress serotonin (MRN and DRN) and to lower dopamine (SNc, VTA) firing during REM. Although serotonin, noreinephrine and histamine neurons cease firing during REM sleep, dopamine neurons show bursting activity in both wakefulness and REM sleep.

Main LHb input is from GPi, LH, VTA, DRN, MRN, LPO (reciprocal), MS/vDBB, SI, NBM, PH, PAG and BNST (Herkenham and Nauta, 1977; Sutherland, 1982). LHb projects to SUM (Kiss et al., 2002), and medial LHb innervates intermediate IPN (Kim, 2009). LHb inhibits dopaminergic VTA/SNc and serotonergic DRN/MRN, and stimulates GABAergic RMTg that directly inhibits DRN, VTA and SNc (Perroti et al., 2005; Jhou et al., 2009 a, b; Kaufling et al., 2009). LHb projects also to LC, TMN, PH, LH, LDT, PPT, NBM, SI, MS/DBB, SUM, LPO, ZI, PVT, MDT (involved in value-based choice selection), raphe pontis nucleus, contralateral LHb and to REM-promoting PnO (Herkenham and Nauta, 1979; Araki et al., 1988). So my prediction is that LHb suppresses the SWS-promoting LPO and IPN. Lateral habenula inhibition of MRN is already well-known. This model proposes that LHb attenuates wake-on cholinergic, GABAergic or glutamatergic neurons in LDT, PPT, NBM, SI, MS/DBB, then noradrenergic LC, histaminergic PH/TMN, and orexinergic LH neurons, glutamatergic SUM. Further prediction is that LHb attenuates also VTg, DTg, ZI, SC, PAG, CM (centromedian thalamic nucleus) and MDT (deselecting choices) to decrease movement and arousal after defeat, after repeatedly bad feedback/outcomes, in chronically hostile environment and during REM sleep. Chronic pain and negative feedback overstimulate LHb, causing learned helplessness, to stop us moving and exerting/losing energy for things that repeatedly hurt and harm us and decrease our well-being or survival.

PPT stimulates STN during REM sleep (Fernández-Mendoza et al., 2009) enhancing the fast (15–35 Hz) subthalamic oscillatory activity. STN then activates LHb by stimulating internal globus pallidus (GPi), as GPi has strong glutamatergic input to LHb. So probably the firing of subthalamic neurons activates GPi, which then causes activation of LHb during REM sleep. This REM (and preREM, shortly before REM) sleep stimulation of LHb helps to suppress dopamine and serotonin release during REM, so prevents switching into waking and into SWS state, respectively. Prolonged LHb activation might lead to suppression of SUM and weakened theta oscillations in the brain, as the LHb projects to SUM, and this model predicts that SUM is suppressed by LHb. In addition the SNr inhibits PPT. It is not clear if the LHb inhibits the wake -on PPT/LDT neurons (which stimulate VTA/SNc) or also the REM-on LDT/PPT neurons. LHb might not suppress REM-on LDT/PPT groups (as they are used to trigger the theta-frequency in SUM during REM sleep), but might attenuate wake/REM-on LDT and PPT neurons leading to decrease in arousal and awareness. If would LHb inhibit REM-on LDT/PPT, it might cause gradual suppression of REM state and support the switching/alternations between the REM sleep and SWS.

The PPT/LDT is known to activate SubC and medullary reticular nuclei that inhibit motor neurons during REM sleep, causing muscle atonia. Overstimulation of LHb, by chronic pain, loss, worries, bad outcomes, or by low basal dopamine or serotonin signal in brain (leading to LHb disinhibition) actually prolongs REM sleep and shortens SWS. The reason for this might be that the shortage of brain serotonin signal when feeling down disinhibits LHb what leads to further suppression of serotonin release. The lack of serotonin then disinhibits the REM-on circuit and shortens SWS stage. Increased REM sleep duration was found in depression (Steriade and McCarley, 1990). This finding supports my prediction that LHb is one of the effectors activated by REM-on neurons in LDT/PPT, possibly via STN.

Interestingly LHb projects to PnO, so might promote (or control) REM sleep not only by suppression of dopamine and serotonin. Model predicts that wake-on orexinergic LH neurons inhibit LHb and RMTg, thus causing rise in VTA/SNc dopamine and reactivity (i.e. when hungry). The REM-on, melanin concentrating hormone (MCH) neurons of LH might inhibit dopaminergic A11 stimulation of motor neurons, or activate SubC or PnO. Further prediction of this model is that LHb suppresses NBM activity, e.g. in depression and after repeated defeat and loss, what leads to weakened attention, working memory strength via weakened fast gamma based cortical coupling. As tonic, fast oscillatory (20-40 Hz) cortical activity is elicited by NBM stimulation (Metherate et al., 1992).

Reticular nucleus pontis oralis and caudalis. Some of their interactions.

Subcoeruleus, nucleus pontis caudalis (PnC) and oralis (PnO/RPO) receive REM-on LDT/PPT input and promote REM sleep. Both PnC and PnO project to oculomotor/visual system. The PnC projections to STN might enhance the REM linked motor response suppression. Movement suppression in REM is caused by SubC that inhibits spinal motor neurons via reticular medullar nuclei. Wake-on norepinephrine is known to inhibit SubC while the wake-on PPT group inhibits PnC. The theta rhythm inducing PnO projects to SUM, LDT, IPN, lateral mamillary body (LMB), mesencephalic reticular formation (that projects to reticular thalamic, reuniens and subthalamic nucleus), retrorubral nucleus, VTA, SNc, zona incerta (ZI), specific PAG regions and CM (Vertes and Martin, 1998). So question is why the PnO interacts with the LMB during REM sleep? The LMB with its head-direction cells receives DTg and postsubicular hippocampal input, and projects to DTg and anterodorsal thalamic nucleus AD. The AD is reciprocally connected with postsubicular hippocampus and with retrosplenial cortex. The LMB might (besides spatial navigation) support navigation on temporal axes (distant past, less distant past, things ahead..) of memory, what is not a strong feature of dreams. Content of dreams is often evolving in incongruous temporal context, with jumps between scenes. Perhaps a random activation of LMB and reuniens during REM serves to mix up and link distant and fresher events and memories into wider cognitive map.

Conclusions

Theta oscillations are evoked either during active waking state by what is going on around us, so by contextual and stimulus based novelty, salient sensory input, interesting or meaningful stimuli, good/valuable or bad/harmful things. Theta rhythm is induced also by REM-on PnO projections to SUM, or via LDT projections to SUM and MS/DBB. In quiet waking without novelty and during consumatory (repetitive) behaviour there is low need to record new information, so it was linked to replay of previously encoded information, after the theta rhythm enabled recording events, contexts and interrelations between things. Possibly, when the hippocampal dentate gyrus gets full and new information lead to interference, the hippocampus stimulates posterior septum to induce SWS via MHb-pathway, to clean up short term memory storage for new input. The MHb → IPN → MRN circuit then promotes SWS and antagonizes the theta promoting circuit, wake and REM sleep. Besides the main idea, this study brought new predictions about interactions of some SWS- versus theta- promoting regions with LHb, RMTg, LPO, VTg and NI. And their effects on sleep/wake control.

This model predicted that lateral habenula projections to SUM have inhibiting effects and attenuate theta rhythm. It also proposed that LHb suppresses the wake-promoting substantia innominata/nucleus basalis (SI/NBM), as well as the SWS-promoting LPO. LHb is known to suppress (for miliseconds) dopamine and serotonin release. Consequent decrease in dopamine input then lowers the NBM output (as dopamine stimulates NBM). One of the purposes of dopamine suppression by LHb might be to decrease arousal after chronic defeat and chronic negative feedback, to stop us moving and losing energy for bad choices, to unlearn wrong (no more valid) ideas, and to deselect the suboptimal choices, decisions and prediction from working memory. Because the LHb attenuates NBM via suppressing dopamine in system, and LHb projects to NBM, the LHb might act in similar way on NBM and directly inhibit it.

Further my circuit-based findings suggest that LHb reciprocally inhibits LPO and orexinergic lateral hypothalamus. While LPO possibly activates RMTg. This study also suggested how can LHb promote REM sleep. Via the REM-on PPT neurons stimulation of STN that activates SNr and GPi, which then activate LHb to suppress dopamine and serotonin release during REM. The IPN → LPO and LPO → MRN connections link together two SWS-promoting systems: proposed MHb → IPN →MRN and LPO. My prediction is that IPN stimulates LPO and LPO stimulates MRN but it has to be tested.

This work seems to be the first circuit based model of the MHb-IPN-MRN circuit function. It extended the known system for control of wakefulness and sleep, by adding the role of MHb and IPN to the known serotonergic MRN system. The MHb increases its activity during anesthesia, and anesthesia is similar to SWS stage. This model combined connectivity references (on regions involved in sleep-wake control and on MHb) and the available (partial) functional findings from the literature. It predicted how do MHb and IPN promote slow wave sleep: by stimulating serotonergic MRN, as well as by inhibiting the opposing theta-promoting circuit, wakefulness and REM sleep regions. It showed few new interactions of theta-promoting regions, known to enable recording of new information in hippocampus in active wake linked to theta state. SWS is known for replay of relationally and spatio-temporally bound informattion in hippocampus. So because the SWS and REM states are involved in different functions, it has sense that there is opposition/inhibitory effect between MHb-IPN-MRN and the theta- promoting circuit. Different neural regions, neuromodulators and states are needed in recording than in replay. Because of proposed circuit based opposition between SWS - promoting and theta- promoting regions, some new relations and interactions between the less known regions of sleep-wake system could be predicted. For example that LHb inhibits IPN and SUM, or that LPO (SWS promoting region) inhibits LHb and VTA but stimulates RMTg to oppose/suppress REM state.

## Abbreviations

DRN dorsal raphe nukleus

DTg dorsal tegmental nucleus of Gudden

GPi globus pallidus pars interna

IPN interpeduncular nukleus

LC locus coeruleaus

LDT lateral dorsal tegmental nukleus

LH lateral hypothalamus

LHb lateral habenula

LPO lateral preoptic nucleus

LS lateral septum

MDT mediodorsal thalamus

MHb medial habenula

MRN median raphe nucleus

MS/DBB medial septum/diagonál band of Broca

NBM nukleus basalis of Meynert

NI nucleus incertus

PH posterior hypotalamus

PPT pedunculopontine nucleus

REM rapid eye movement

RMTg rotromedial tegmental nukleus

SI substantia innominata

SNc Substantia nigra compacta

SNr Substantia nigra reticulata

STN subthalamic nukleus

SUM supramamillary nukleus

SWS slow wave sleep

TMN tuberomamillary nucleus

VTA vetral tegmental nucleus

VTg ventral tegmental nucleus of Gudden